\documentclass[peerreview]{IEEEtran}
\renewcommand{\baselinestretch}{1.5}
\usepackage{amsmath,amssymb}

\usepackage{amssymb}
\usepackage{bm}
\usepackage[dvipdfmx]{graphicx}
\usepackage{color}

\graphicspath{{figure/}}

\newcommand{\defeq}{\stackrel{\triangle}{=}}

\usepackage{setspace}

\begin{document}
\title{Performance Analysis based on Density Evolution \\ on Fault Erasure Belief Propagation Decoder} 
\author{
  \IEEEauthorblockN{Hiroki Mori and Tadashi Wadayama} \\
  \IEEEauthorblockA{Nagoya Institute of Technology,   Japan\\
      Email: mori@it.cs.nitech.ac.jp, wadayama@nitech.ac.jp} 
}

\maketitle
\begin{abstract}
In this paper, we will present an analysis on the fault erasure BP decoders
based on the density evolution.
In the fault BP decoder, messages exchanged in a  BP process 
are stochastically corrupted due to unreliable logic gates and flip-flops;
i.e., we assume circuit components with transient faults.
We derived a set of the density evolution equations for the fault erasure BP processes.
Our density evolution analysis reveals the asymptotic behaviors of the estimation error probability 
of the fault erasure BP decoders. In contrast to the fault free cases, it is observed that 
the error probabilities of the fault erasure BP decoder converge to positive values, and 
that there exists a discontinuity in an error curve corresponding to the fault BP threshold.
It is also shown that an message encoding technique provides 
higher fault BP thresholds than those of the original decoders at the cost of increased circuit size.
\end{abstract}

\section{Introduction}\label{sec:intro}

Recent advance of CMOS technology leads to denser VLSI implementation and 
this trend is continuing \cite{ITRS}．
In near future, faulty behaviors of logic gates and flip-flops due to cosmic rays 
or thermal noises would become more problematic \cite{fault1}．
We should take care of fault tolerant VLSI design to attain 
highly reliable circuits based on unreliable components \cite{fault-tolerant1}\cite{fault-tolerant2}．

In this paper, we call a decoder for an error/erasure correcting code (ECC) composed by unreliable components 
a {\em fault decoder}.  Fault tolerance of the decoder is of critical importance 
because ECC is often exploited for ensuring high reliability of data memories in a circuit.
Therefore, in a digital system based on unreliable components, 
ECC behaves as a key component to compose reliable circuits.
Another reason for studies on fault decoders comes from the {\em packet-based communication 
in a VLSI chip.}
A new paradigm of data exchange in CPU,  {\em Network on Chip} (NoC), 
is actively studied for replacing conventional on-chip buses for data/address 
exchange in a chip \cite{network-on-chip}.
An NoC system is based on a packet-based network connecting many CPU cores and routers 
for packet switching. If the network is congested, packet erasures due to collisions at a router 
may occur and compensation for erased packets is needed．Erasure correction would be
a one of solutions for such packet erasures in a chip \cite{LT}.

Several works discussing fault decoders for {\em Low-Density Parity-Check} (LDPC) codes have been published.
In 2011, Varshey presented an analysis for the fault Gallager-A decoder \cite{fault-decoder1}．
He assumed a probabilistic  model such that independent 
transient faults may occur in a circuit of the Gallager-A decoder.
A fault causes deterioration of the quality of the messages exchanged in a decoder and 
it results in degradation of the decoding performance.
Based on these assumptions, analysis based on the density evolution  was presented in \cite{fault-decoder1}.
Sadegh et. al showed a similar analysis on the fault Gallaber-B decoder \cite{fault-decoder2}．
They also derived the density evolution equations for the fault Gallaber-B decoder
and calculated the thresholds for $q$-ary symmetric channel.
Other related works on the fault decoders can be found in 
\cite{fault-decoder3}\cite{fault-decoder4}\cite{fault-decoder5}\cite{fault-decoder6}.

A goal of this work is to analyze the asymptotic behavior of the {\em fault erasure belief propagation (BP) decoder}
based on the density evolution. It is expected that the results obtained for fault erasure BP decoder 
give us a useful insight for appropriate design of BP decoders made from unreliable components.

\section{Fault erasure BP decoder}\label{sec:modeling}
A fault erasure BP decoder is a BP decoder for memoryless erasure channels
based on unreliable components such as logic gates and flip-flops.
In this section, we are going to define a fault erasure BP decoder.

\subsection{Fault model for erasure BP decoder}

In this paper, we assume independent transient faults of logic gates and flip-flops and 
do not assume occurrences of the permanent faults.
The occurrence of transient faults are modeled by a probabilistic model.
Namely, transient faults are assumed to be independent events and 
the probability of occurrences of the fault does not depend on the places. 
This model is based on the Neumann model \cite{Neumman-model} and 
it was used in the related literatures \cite{fault-decoder1} \cite{fault-decoder2}.

\begin{figure}[tbp]
  \begin{center}
    \includegraphics[scale=0.30]{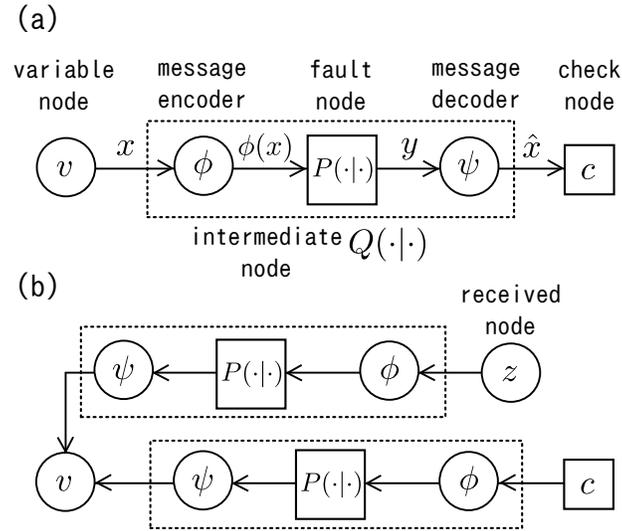}
  \end{center}
  \caption
{
A message flow in erasure BP process: (a) variable to check message flow, (b) check to variable message flow.
}
\label{fig:faulty_node}
\end{figure}

In order to clarify the definition of the fault model used in the paper,  
we focus on an erasure correction BP process. 
Figure \ref{fig:faulty_node} presents 
a message flow from a variable node $v$ to a check node $c$ in a Tanner graph.
Three nodes, called {\em message encoder}, {\em fault node}, 
and {\em message decoder}, are inserted in between the variable and check nodes.
The message encoder encodes a BP message in the message alphabet $\{0,1, e\}$ into 
a binary (i.e., $\{0,1\}$) sequence 
that are stored in flip-flops. The message decoder 
estimates a BP message in $\{0,1, e\}$ from a given binary sequence that is the read-out 
symbols from the flip-flops.  The precise definition of 
the pair of an encoder and a decoder will be given later. 
We assume that a binary symbol stored in a flip-flop can be flipped with 
probability $\alpha (0 \le \alpha < 1)$ due to independent transient faults.
The fault node in  Fig. \ref{fig:faulty_node} corresponds to 
the memoryless binary symmetric channel with the bit-flip probability $\alpha$. 

According to Fig.\ref{fig:faulty_node} (a), we will explain the details of the message encoding and the probabilistic 
model for transient faults.
The message of a BP process  is expressed with the message alphabet $\{0,1,e\}$ where $e$ represents an erasure.
The variable node $v$ encodes a message  into a binary sequence of length 2 that 
is suitable for storing in a 2 flip-flops.
The {\em message encoding function} $\phi: \{0,1,e\} \rightarrow \{0,1\}^2$ is defined by
\begin{eqnarray}\label{eqn:message_coding}
\phi(x) \defeq \left\{\begin{array}{ll}
00, & x=0, \\
11, & x=1, \\
01, & x=e.\\
\end{array} \right.
\end{eqnarray} 
The output of the message encoder (two binary symbols) are stored in a pair of flip-flops.

The transient faults are modeled by probabilistic bit flips.
A binary information in a flip-flop may alter its value with probability $\alpha$ and 
this bit flip events are independent.
Thus, the conditional probability $P(m' | m) (m' \in \{0,1\}^2, m  \in \{0,1\}^2)$ is given by
\[
P(m'|m) = (1 - \alpha)^{2 - d_H(m',m)} \alpha^{d_H(m',m)}
\]
where $d_H$ represents the Hamming distance. The symbol $m$ and $m'$ denote 
bit sequences of length 2 stored in the flip-flops.
The message decoder tries to estimate a message sent from the variable node $v$
from the read-out symbols from the flip-flops $y\in\{0,1\}^2$.
The decoding function $\psi: \{0,1\}^2 \rightarrow \{0,1,e\}$  is given by 
\begin{eqnarray}\label{eqn:message_decoding_function}
\psi(y) \defeq \left\{\begin{array}{ll}
0, & y = 00, \\
1, & y = 11, \\
e, & y \in \{01,10 \}. \\
\end{array} \right.
\end{eqnarray}
Finally, the check node $c$ obtains the estimate of a message $\hat x = \psi(y)$.
In the following analysis, it is convenient to derive the conditional probability 
of $\hat x$ given $x$, which is denoted by $Q(\hat x| x)$.
From the definitions of the message encoding and the probabilistic model 
for the transient faults,  the conditional probability can be immediately derived as 
\begin{eqnarray}\nonumber
\left(
\small
\begin{array}{ccc}
Q(0|0)&Q(1|0)&Q(e|0)\\
Q(0|1)&Q(1|1)&Q(e|1)\\
Q(0|e)&Q(1|e)&Q(e|e)
\end{array}
\right)
=\\\label{eqn:message_change_matrix}
\left(
\small
\begin{array}{ccc}
(1-\alpha)^2&\alpha^2&2\alpha(1-\alpha)\\
\alpha^2&(1-\alpha)^2&2\alpha(1-\alpha)\\
\alpha(1-\alpha)&\alpha(1-\alpha)&\alpha^2+(1-\alpha)^2
\end{array}
\right).
\end{eqnarray}

Figure \ref{fig:faulty_node}(b) indicates a message flow in the reverse direction.
It includes a node $z$ representing a received symbol.  In this case, the same 
encoding function,  the decoding function, and the probabilistic fault model are assumed.
The dashed box in Fig. \ref{fig:faulty_node} (a)(b) corresponding to 
this conditional probability $Q(\hat x| x)$ is also called an {\em intermediate node} in a block diagram.

\subsection{Modification of variable node operation}

In a conventional erasure BP process, 
there is no possibility for a variable node to receive 
contradicting input messages from adjacent check nodes simultaneously.
However, in a fault erasure BP process defined above, 
a variable node may have messages containing both 0 and 1 simultaneously.
We thus need to modify the variable node process for accepting such contradicting messages.
In this paper, we adopt the following simple modification on the variable node process. 
If a variable node receives 
a set of contracting messages that include both 0 and 1,  then the variable node 
sends the erasure symbol to the neighboring check nodes.
The same rule is applied to the process for determination of the estimate symbol.

\section{Density evolution equations}\label{sec:DE}

The density evolution (DE) is an important method to unveil the asymptotic behavior of a BP decoding algorithm.
In a DE process, we can track the time evolution of
the probability distribution of messages  (or the probability density function 
in a case where the messages are continuous).
The asymptotic probability distributions obtained by iterative computation
tell us the asymptotic quantitative features of the decoding algorithm.
In this section, we will derive the DE equations for the fault erasure BP decoder.

\subsection{Derivation of DE equations}

In the following analysis, we will make several assumptions that have been commonly used 
in related works.
The channel is assumed to be a memoryless binary erasure channel (BEC) with 
the erasure probability $\epsilon (0 \le \epsilon \le 1)$.
In this paper,  we consider a regular LDPC code ensemble with 
the variable node degree $d_v$ and the check node degree $d_c$.
The transmitted word is assumed to be the zero codeword of infinite length.

Suppose that $x$ represents an input to an intermediate node 
(corresponding to the conditional probability $Q(\cdot|\cdot)$)
and that $\hat x$ represents the corresponding output from the intermediate node.
If $x$ is distributed according to the probability distribution $t(\cdot)$
over the message alphabet $\{0,1,e \}$, then the probability distribution of the output $\hat x$ 
obeys $t'(\cdot)$ given by
\begin{equation}\label{inter}
t'(\hat x) =  \sum_{x \in \{0,1,e\}} Q(\hat x| x) t(x).
\end{equation}

\begin{figure}[tbp]
  \begin{center}
    \includegraphics[scale=0.3]{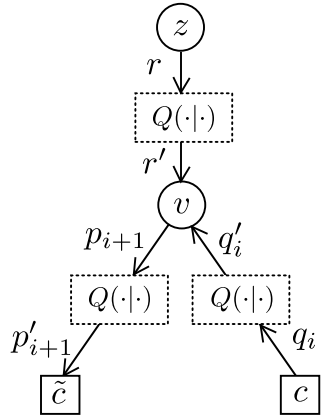}
  \end{center}
  \caption
{
Relation of message probability distributions
}
  \label{fig:DE}
\end{figure}

In the following, the details of the probability distributions are introduced according to Fig.\ref{fig:DE}.
The probability distribution corresponding to the message 
emitted from a check node $c$ is denote by $q_i$. 
The index $i$ represents the discrete time index in an iterative process.
A message from a check node enters an intermediate node, which represents 
the effect of the probabilistic faults.
The distribution corresponds to the output of the intermediate node 
is represented by $q'_i$ that is given by
\begin{eqnarray}\label{eqn:q'}
q'_i (\hat x) = \sum_{x \in \{0,1,e\}} Q(\hat x| x) q_i(x).
\end{eqnarray}
In the following, we will use a convention such that the symbol for the output distribution of 
the intermediate node is expressed with the symbol of the input distribution with the prime symbol
such as $t$ and $t'$.

A variable node $v$ computes a message from the set of messages it receives.
The message distribution corresponding to the message from a variable node to 
an intermediate node follows the distribution $p_{i+1}$.
The corresponding output distribution from the intermediate node is given by
\begin{eqnarray}\label{eqn:p'}
p'_{i+1} (\hat x) = \sum_{x \in \{0,1,e\}} Q(\hat x| x) p_{i+1}(x).
\end{eqnarray}

We also assume that probabilistic faults may occur 
in a message flow from a received symbol node to a variable node.
A received symbol node send a message in $\{0,1,e\}$ to an intermediate node
according to its received value.
The probability distribution $r(\cdot)$ of the message is given by
\begin{eqnarray}\label{eqn:r}
r(x) \defeq \left\{\begin{array}{ll}
1-\epsilon, & x = 0, \\
0, & x =1, \\
\epsilon, & x = e. \\
\end{array} \right.
\end{eqnarray}
The message distribution of the corresponding message from the intermediate node follows
the distribution 
\begin{eqnarray}\label{eqn:r'}
r' (\hat x) = \sum_{x \in \{0,1,e\}} Q(\hat x| x) r(x).
\end{eqnarray}

We first derive the DE equations on the check node output.
There are two cases depending on the output of the check node.
Firstly, consider the case where the output of the check node is 0.
The check node $c$ calculates a message to an adjacent variable node $v$.
If and only if the set of $d_c-1$ incoming messages received by $c$ except for the one from $v$ 
contain even number of 1's and contains no erasure symbols, 
then the message from $c$ becomes 0.
Therefore, the distribution of the check node output $q_i(0)$ is given by
\begin{eqnarray}\nonumber
q_i(0)&\!=\!&\sum_{j: even }^{d_c-1}\binom{d_c-1}{j}p'_i(1)^{j}p'_i(0)^{d_c-1-j}\\
&\!=\!&\frac{(p'_i(0)\!+\!p'_i(1))^{d_c-1}}{2}\!+\!\frac{{(p'_i(0)\!-\!p'_i(1))}^{d_c-1}}{2}. \label{eqn:q0}
\end{eqnarray}
The binomial theorem is used in the derivation above.
In a similar manner,  we can derive $q_i(1)$.
Note that, in this case, the set of the $d_c-1$ messages consisting of odd number of 1's and no erasure symbols
leads to the output message 1 from the check node. We thus have 
\begin{eqnarray}\nonumber
q_i(1)&\!=\!&\sum_{j: odd}^{d_c-1}\binom{d_c-1}{j}p'_i(1)^{j}p'_i(0)^{d_c-1-j}\\
&\!=\!&\frac{(p'_i(0)\!+\!p'_i(1))^{d_c-1}}{2}\!-\!\frac{{(p'_i(0)\!-\!p'_i(1))}^{d_c-1}}{2}. \label{eqn:q1}
\end{eqnarray}

We will then consider the DE equations on the variable node output.
Let us assume that an output of a variable node is 0, and
that the variable node $v$ calculates a message to an adjacent check node $c$.
Let $M$ be  the set of the $d_v-1$ incoming messages to $v$ from adjacent check nodes 
except for the one from $c$.
The variable node message becomes 0 if and only if 
the event (A) $y=0$ and $1 \notin M$ holds,  or  the event (B) $y=e$, $1 \notin M$, and $0 \in M$ 
holds where $y$ received symbols corresponding to the variable node $v$.

The probability corresponding to the event (A) becomes
$
Prob[A] = r'(0) (1 - q_i'(1))^{d_v-1}
$
because all the incoming messages are independent.
The probability of the event (B) is given by 
$
Prob[B] = r'(e) \left(  (1-q'_i(1))^{d_v-1}-q'_i(e)^{d_v-1} \right).
$
Since these two events are independent, the probability $p_{i+1}(0)$
is the sum of these two probabilities:
\begin{eqnarray} \nonumber
p_{i+1}(0)&=& Prob[A] + Prob[B] \\ \nonumber
&=& r'(0)(1-q'_i(1))^{d_v-1} \\
&+& r'(e) \left(  (1-q'_i(1))^{d_v-1}-q'_i(e)^{d_v-1} \right). \label{eqn:p0}
\end{eqnarray}
In a similar manner, we can derive the probability corresponding to the variable node message to be $1$:
\begin{eqnarray} \nonumber
p_{i+1}(1)
&=&r'(1)(1-q'_i(0))^{d_v-1} \\
&+&r'(e) \left( (1-q'_i(0))^{d_v-1}-q'_i(e)^{d_v-1} \right). \label{eqn:p1}
\end{eqnarray}

It should be remarked that, for any discrete time index $i$, 
the equalities
$
p_i(0) + p_i(1) + p_i(e) = 1
$
and 
$
q_i(0) + q_i(1) + q_i(e) = 1
$
hold.

From the arguments above, we have all the DE equations required for the DE analysis of
the fault erasure BP decoding. Namely, Based on 
Eqs. (\ref{inter})(\ref{eqn:r})(\ref{eqn:q0})(\ref{eqn:q1})(\ref{eqn:p0})(\ref{eqn:p1})
with the initial condition $q_0(0)=0, q_0(1)=0$,
an iterative calculation on the message probability functions leads
to the asymptotic message distributions.

\subsection{Asymptotic error probability}

According to the conventional erasure BP rule, 
if incoming messages to a variable node contain no 1's and contain a 0,
then the tentative estimate of the variable node becomes 0.
Let us denote the probability for such an event by $s_{i}(0)$.
The probability $s_{i}(0)$ is given by
\begin{equation}
s_{i}(0)=r'(0)(1-q'_i(1))^{d_v}+r'(e) \left((1-q'_i(1))^{d_v}-q'_i(e)^{d_v}  \right).
\end{equation} 
A DE process can evaluate 
the asymptotic error probability 
\begin{equation}
\gamma(\epsilon, \alpha) \defeq \lim_{i \to \infty}(1-s_i(0)).
\end{equation}
In the following parts of this paper, we will focus on the behavior 
of the asymptotic error probability $\gamma(\epsilon, \alpha)$.

\section{Numerical results}\label{sec:result1}

In the previous section, we derived the DE equations for the fault erasure BP decoder.
In this section, numerical results indicating the asymptotic behavior of the decoder 
will presented. 

\subsection{Effect of transient faults}
\begin{figure}[tbp]
  \begin{center}
    \includegraphics[width=8cm,height=6cm]{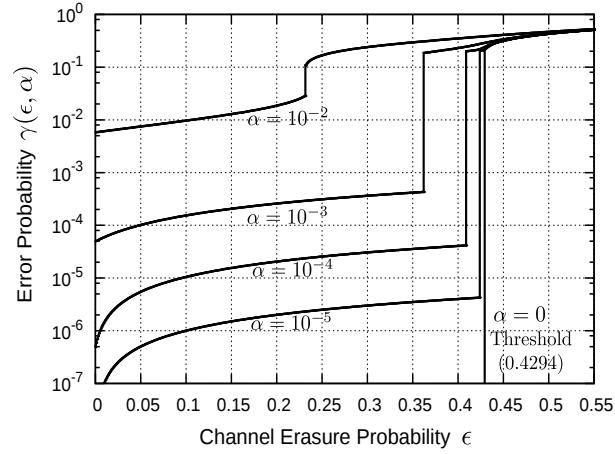}
  \end{center}
  \caption
{
Relationship between erasure probability $\epsilon$ and asymptotic error probability $\gamma(\epsilon,\alpha)$ 
((3,6)-regular LDPC code ensemble)
}
  \label{fig:result1}
\end{figure}

Figure \ref{fig:result1} presents the asymptotic error probabilities $\gamma(\epsilon,\alpha)$
for $(d_v,d_c)=(3,6)$-regular LDPC code ensemble.
The four curves depicted in Fig.\ref{fig:result1} correspond  to 
the fault probabilities $\alpha=10^{-2},10^{-3},10^{-4},10^{-5}$  from left to right.
When the fault probability $\alpha$ is equal to 0, the system model exactly coincides with the 
common erasure BP decoder model without transient faults. In such a case, the asymptotic error probability 
converges to 0 if $\epsilon$ is greater than the BP threshold $\epsilon_{BP}=0.42944$ 
(this value is also included in Fig. \ref{fig:result1}).
In the case of positive $\alpha$, the situations are totally different.
When $\alpha>0$, we can observe that
$\gamma(\epsilon,\alpha)$ converges to positive values due to the faults occurred in the BP decoder.
From this figure, it is also seen that smaller $\alpha$ gives smaller $\gamma(\epsilon,\alpha)$.
Furthermore, each curve has a sudden (vertical) jump at a certain erasure probability.
For example, the curve of $\alpha=10^{-3}$ shows $\gamma(\epsilon,10^{-3}) > 10^{-1}$ 
in the regime $\epsilon > 0.36207$.  On the other hand, in the regime $\epsilon < 0.36207$,
$\gamma(\epsilon,10^{-3})$ takes the values smaller than $10^{-3}$. The behaviors of the decoder 
are sharply separated at the erasure probability $\epsilon = 0.36207$ that is considered to be 
a threshold value for the fault erasure BP decoder.

\subsection{BP dynamics of fault erasure BP decoder}
\begin{figure}[t]
  \begin{center}
    \includegraphics[width=8cm,height=6cm]{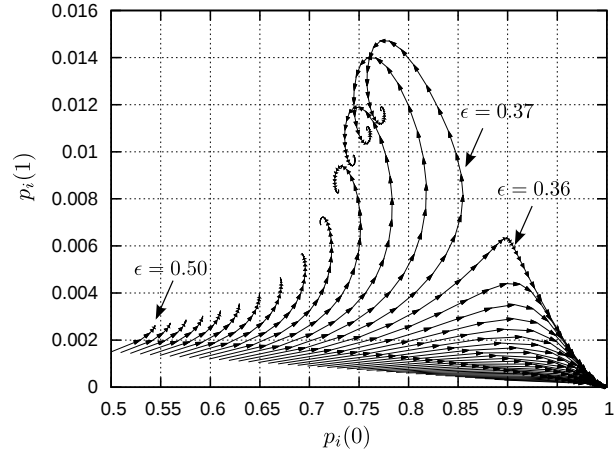}
  \end{center}
  \caption
{
Dynamics of BP process: Evolution of message probabilities for variable nodes $p_i(0)$ and $p_i(1)$
($(3,6)$-regular LDPC code ensemble，fault probability $\alpha=10^{-3}$)
}
  \label{fig:result2}
\end{figure}

Figure \ref{fig:result2} indicates the dynamics of the BP processes via
the evolutions of the pair of the message probabilities  $p_i(0)$ and $p_i(1)$.
The ensemble is $(3,6)$-regular LDPC code ensemble and the fault probability is assumed 
to be $\alpha=10^{-3}$. Each arrow in the figure shows a change of the message probabilities from 
$(p_i(0),p_i(1))$ to  $(p_{i+1}(0),p_{i+1}(1))$ and 
each trajectory corresponds to an erasure probability in the range $\epsilon = 0.01 j (10 \le j \le 50)$.
Since the zero codeword  is assumed to be transmitted,  the probability $p_i(0)$ 
represents the probability for the correct decoding.  
It is immediately recognized that there are two groups of the trajectories: one group 
corresponds to the range $0.37 \le \epsilon \le 0.5$ and the other group corresponds to the range
$0.1 \le \epsilon \le 0.36$. The trajectories in the first group show the upward movements. 
This means that the error probability tends to converge to a higher value.
On the other hand, the trajectories in the second group indicate that $p_i(0)$ approaches 
to 1 as the number of iterations increases. This numerical results strongly suggest 
the existence of  a {\em bifurcation} of this DE evolution processes 
that can be considered as a non-linear dynamical system. 
At the erasure probability that corresponds to this bifurcation, we  can observe sudden drop
of the asymptotic error probability in Fig.\ref{fig:result1}.

\subsection{Degree and asymptotic error probability}

\begin{figure}[t]
  \centering
  \includegraphics[scale=1.1]{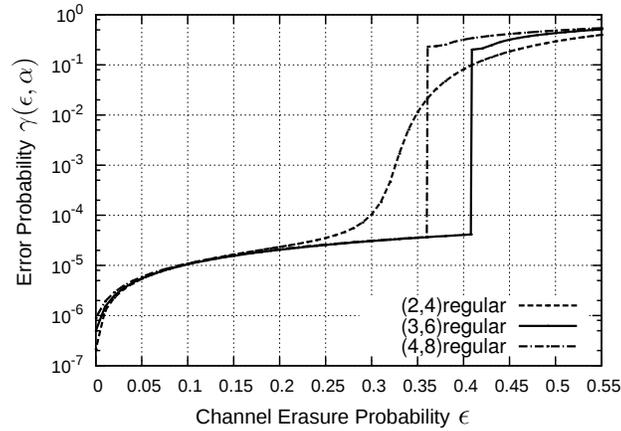}
  \caption{
Relationship between erasure probability $\epsilon$ and asymptotic error probability $\gamma(\epsilon,\alpha)$
($(d_v,d_c)$-regular LDPC code ensemble，design code rate 0.5，fault probability $\alpha=10^{-4}$)
}
\label{fig:change_degree}
 \end{figure}

Figure \ref{fig:change_degree} presents the asymptotic error probabilities $\gamma(\epsilon,\alpha)$
for regular-LDPC code ensembles with degrees $(d_v,d_c)=(2,4),(3,6),(4,8)$.
All the ensembles correspond to the design code rate $1/2$. The fault probability is set to $\alpha=10^{-4}$.
From Fig.\ref{fig:change_degree}, we can observe that  the $(d_v,d_c)=(3,6)$ ensemble 
provides the highest fault BP threshold.
It is well known that $(3,6)$ ensemble gives the highest threshold in the fault free cases.
Similar tendency can be seen in the cases where transient faults exist.

\section{Message encoding}\label{sec:message_encoding}

In the previous section, we observed numerical results of the DE analysis on the 
fault erasure BP decoder. According to the numerical results, it was shown that
we must admit non-zero error probability even in the asymptotic regime if the fault probability is positive.
In this section, we discuss and compare two methods for improving the decoding performance 
of the fault erasure BP decoding at the cost of increased hardware complexity.

The simplest way to improve the decoding performance is to exploit several 
identical erasure BP decoders in parallel. 
By using the majority votes from the outputs obtained from these component decoders,
we can obtain more reliable estimates of transmitted symbols.
In this paper, the scheme is called {\em Majority voting scheme}.
Another way to enhance the reliability is to use a longer code to protect BP messages.
In Section 2, we introduced the message function $\phi$ that encodes a BP message 
into 2-binary symbols.  By replacing the encoding function to an encoding function for a longer code,
we can expect that the immunity against possible faults becomes stronger.
We call this scheme {\em message encoding}.
Of course, both schemes (i.e., majority voting and message encoding) require
increase of the circuit size  that can be considered as the cost should be paid for the improvement of the immunity.

\subsection{Majority voting scheme}

In this subsection, we introduce a simple majority voting scheme that 
emploies $N$-fault erasure BP decoders for  improving the fault immunity.
The majority voting scheme determines its output by majority voting 
based on $N$-outputs from the component BP decoders.
Although this scheme requires $N$-fold circuit size compared with
the single fault erasure BP decoder, it is expected that the majority voting process 
improves the asymptotic error probability.

In the following, we will discuss the case where $N=2$. 
The argument below can be easily extended to general cases where $N>2$.
Let $\hat{x}^{(1)},\hat{x}^{(2)}\in\{0,1,e\}$ be the decoder outputs from 
the two component decoders $D^{(1)}$ and $D^{(2)}$. 
The majority voting process is defined the function
\[
\tau(\hat{x}^{(1)},\hat{x}^{(2)})=\left\{ \begin{array}{ll}
0,&(\hat{x}^{(1)},\hat{x}^{(2)})=(0,0),(0,e),(e,0)\\
e,&(\hat{x}^{(1)},\hat{x}^{(2)})=(0,1),(1,0),(e,e)\\
1,&(\hat{x}^{(1)},\hat{x}^{(2)})=(1,1),(1,e),(e,1),\\
\end{array} \right.
\]
where the function $\tau$ represents the output from the majority voting decoder.
We denote the asymptotic error probability for the majority voting scheme 
by  $\gamma_{maj}(\epsilon,\alpha)$.

In the following, a lower bound on $\gamma_{maj}(\epsilon,\alpha)$ will be discussed.
Throughout the following argument, we assume that 
\begin{eqnarray}\label{eqn:maj_theorem_assumption}
s(0,0)&\geq& (s(0))^2 
\end{eqnarray}
holds where the quantity $s(0)$  in the righthand side  is the asymptotic value of $s_0(x)$; i.e., 
$
s(0) = \lim_{i \rightarrow \infty} s_i(0),
$
which can be evaluated by the density evolution. The quantity $s(0,0)$ is the asymptotic joint probability 
corresponds to the event that two decoder outputs take the value $(0,0)$.
This is a natural assumption because the outputs from the the component decoders $D^{(1)},D^{(2)}$ 
are expected to be highly correlated. 
Under the assumption of (\ref{eqn:maj_theorem_assumption}),
we can easily derive a lower bound  of $\gamma_{maj}(\epsilon,\alpha)$
\begin{eqnarray}\label{eqn:parallel_LB}
    \gamma_{maj}(\epsilon,\alpha)\geq(1-s(0))^2.
\end{eqnarray}


\subsection{Details of message encoding}

In this subsection, we will introduce a simple message encoding scheme 
based on a binary code of length $n$. The parameter $n$ is referred to as
{\em message code length}.
In the following, we redefine the encoding and decoding functions.
The encoding function $\phi: \{0,1,e\} \rightarrow \{0,1\}^n$ is an encoding function 
now defined by
\begin{eqnarray}
\phi(x)=\left\{\begin{array}{ll}
\overbrace{00\cdots0}^{n}, & x=0, \\
\overbrace{11\cdots1}^{n}, & x=1, \\
\overbrace{00\cdots0}^{n/2}\overbrace{11\cdots1}^{n/2}, & x=e.\\
\end{array} \right.
\end{eqnarray} 
There are several possibilities for choosing decoding functions corresponding to 
the encoding function defined above.
One simple choice is to define a decoding function $\phi$ as
\begin{eqnarray}
\psi(y)=\left\{\begin{array}{ll}
0, & w_H(y) = 0 \\
1, & w_H(y) = n\\
e, & \mbox{otherwise}, \\
\end{array} \right.
\end{eqnarray}
where $w_H$ represents the Hamming weight function.
In this case, the conditional probability $Q(\hat{x}|x)$ corresponding to 
the intermediate node is given by 
\[
\left(
\begin{array}{ccc}
(1-\alpha)^n&\alpha^n&1-(1-\alpha)^n-\alpha^n\\
\alpha^n&(1-\alpha)^n&1-(1-\alpha)^n-\alpha^n\\
\alpha^\frac{n}{2}(1-\alpha)^\frac{n}{2}&\alpha^\frac{n}{2}(1-\alpha)^\frac{n}{2}&1-2\alpha^\frac{n}{2}(1-\alpha)^\frac{n}{2}\\
\end{array}
\right).
\]
Plugging this condition probability into the DE equations, 
we can evaluate the asymptotic error probabilities.

\subsection{Asymptotic error probabilities for message encoding}\label{sec:result2}

\begin{figure}[t]
  \begin{center}
    \includegraphics[width=8cm,height=6cm]{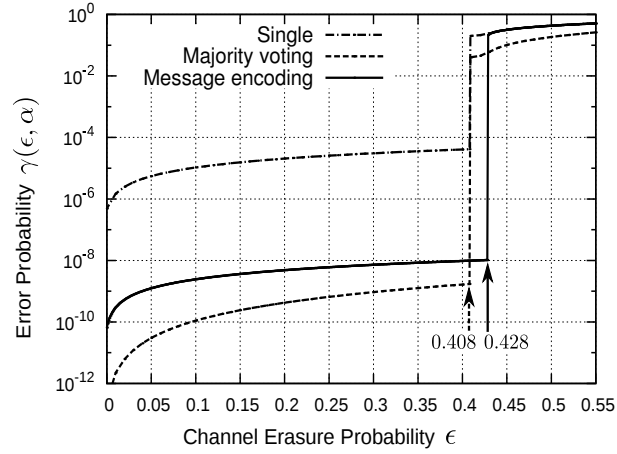}
  \end{center}
  \caption
{
Relationship between erasure probability $\epsilon$ and asymptotic error probability 
$\gamma(\epsilon,\alpha)$($(3,6)$-regular LDPC code ensemble,  fault probability $\alpha=10^{-4}$，
message length in message encoding $n=4$，the number of component decoders in majority logic decoder $N=2$)
}
  \label{fig:result4}
\end{figure}

Figure \ref{fig:result4} presents the asymptotic error probabilities  of
the majority voting decoder (two decoders in parallel, $N=2$) and a fault erasure BP decoder with 
message encoding $(n=4)$. The $(3,6)$-regular LDPC code ensemble is assumed and
the fault probability is set to $\alpha=10^{-4}$.
Both schemes can be considered to have comparable circuit sizes.
In Fig. \ref{fig:result4}, the curve of the majority voting decoder 
corresponds to the lower bound (\ref{eqn:parallel_LB}).
From Fig.\ref{fig:result4}, we can observe that the BP decoder with message encoding archives 
a higher threshold that those of the single BP decoder and the majority logic decoder.
This observation implies that the message encoding has a potential advantage over
the majority logic decoder in terms of the decoding performance close to the threshold.

\subsection{Choice of message decoding function $\psi$}

In a design of an appropriate message encoding scheme,  a choice of message decoding function is critical.
When $n$ becomes large, we have freedom to choose a message decoding function.
In this subsection, we will discuss choices for  a message decoding function.

\begin{figure}[t]
  \centering
  \includegraphics[scale=1.1]{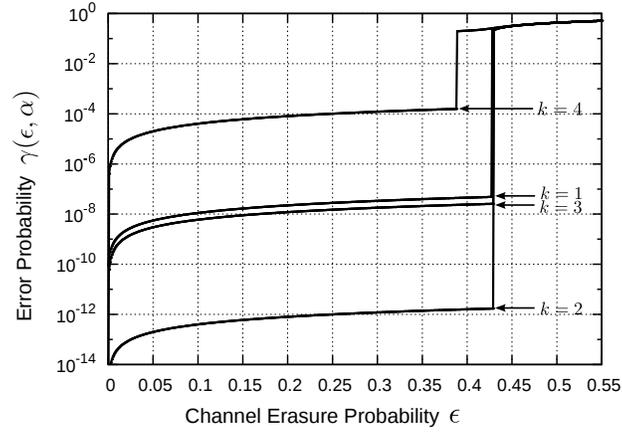}
  \caption{
Relationship between  erasure probability $\epsilon$ and asymptotic error probability $\gamma(\epsilon,\alpha)$
($(3,6)$-regular LDPC code ensemble，fault probability $\alpha=10^{-4}$，
code length $n=8$)
}
  \label{fig:n=8_change_decoding_function}
 \end{figure}

We redefine the message decoding function as 
\begin{eqnarray}
\psi(y)\triangleq\left\{\begin{array}{ll}
0, & 0\leq w_H(y)\leq k-1,\\
1, & n-k+1\leq w_H(y)\leq n,\\
e, & \mbox{otherwise}. \\
\end{array} \right.
\end{eqnarray}
The parameter $k(1\leq k\leq n/2,k\in\mathbb{N})$ controls  the decision region for the messages $\{0,1,e \}$.
For example, As $k$ gets large, the decision region of $e$ become narrower.
Figure \ref{fig:n=8_change_decoding_function} presents 
relationships between the parameter $k$ and 
the asymptotic error probability $\gamma(\epsilon,\alpha)$.
The message code length is assumed to be $n=8$ and the fault probability is set to $\alpha=10^{-4}$.
From Fig.\ref{fig:n=8_change_decoding_function},
we can see that the asymptotic erasure probabilities depends on the parameter $k$.
In this setting, the worst case is $k=4$  and the best case is $k=2$.
This result implies that appropriate choice of the message decoding function 
is important to attain better asymptotic BP decoding performance.

\subsection{Relationship between code length and asymptotic error probability}

\begin{figure}[t]
  \centering
  \includegraphics[scale=1.1]{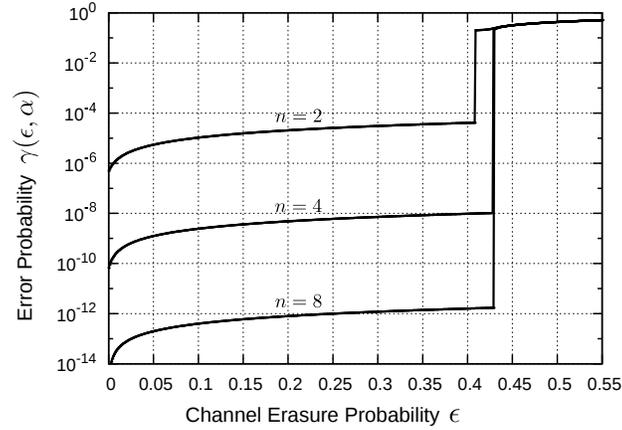}
  \caption
{
Relationship between erasure probability $\epsilon$ and asymptotic error probability $\gamma(\epsilon,\alpha)$.
($(3,6)$-regular LDPC code ensemble, fault probability $\alpha=10^{-4}$，
code length $n=2,4,8$)
}
  \label{fig:change_message_length}
 \end{figure}

Figure \ref{fig:change_message_length} presents the asymptotic error probabilities
for message code length $n=2,4,8$.
Note that we used the optimum parameter for each case such as 
$k=1 (n=2)$, $k=1 (n=4)$, and $k=2(n=8)$.
The regular LDPC code ensemble with $(d_v,d_c)=(3,6)$ is assumed and the fault probability is set to $\alpha=10^{-4}$.
From Fig.\ref{fig:change_message_length}, it is observed that
the asymptotic error probabilities decreases as code length $n$ increases.
However, comparing two cases $n=4$  and $n=8$, we can obtain only small improvement 
in terms of the fault BP threshold.  This means that the major benefit of longer message codes
is lowering the error floor of the asymptotic error probability when $n$ is sufficiently large.

 \section{Conclusion}
 
In this paper, we proposed a model for the fault erasure BP decoders with  transient faults.
Based on the model, the DE equations were derived and used for 
numerical evaluation. The DE analysis shows the asymptotic behaviors of 
the fault erasure BP decoder.
The most notable result revealed via the DE analysis is that 
the asymptotic error probability converges to a positive value in contrast to the the fault free case.
The sudden drop of error probability at a certain erasure probability is considered to be a consequence of
a bifurcation of the DE dynamical system.
In order to improve the decoding performance, we presented two schemes: the message encoding scheme and
the majority voting scheme. The result of the DE analysis indicates that 
the message encoding scheme has clear advantage over the majority voting scheme in terms of 
the fault erasure BP threshold.

\section*{Acknowledgment}

 This work was supported by JSPS Grant-in-Aid for Scientific Research
 (B) Grant Number 25289114.

\end{document}